\documentstyle[epsf,12pt]{article}
\font \msam=msam10
\newcommand{\filledboxes}{{\msam \char 004}}
\newcommand{\filleddiamonds}{{\msam \char 007}}
\newcommand{\filledtriup}{{\msam \char 78}}
\newcommand{\filledtridn}{{\msam \char 72}}
\newcommand{\filledcircles}{\unitlength=1mm\begin{picture}(2,4)
\put(1.419,1.43){\circle*{1.5}}
\end{picture}\ }
\large
\begin{document}

\title{$A$--Dependence of $\Lambda\Lambda$  bond
energies in double--$\Lambda$ hypernuclei}
\author{D.E. Lanskoy, Yu.A. Lurie, and
Andrey M. Shirokov\thanks{E-mail: shirokov@anna19.npi.msu.su} }
\date {Institute of Nuclear Physics, Moscow State University,\\
    Moscow, 119899, Russia }
\maketitle

\begin{abstract}The $A$-dependence of the bond energy $\Delta
B_{\Lambda\Lambda}$ of the
${\Lambda\Lambda}$ hypernuclear ground states is calculated in a three-body
${\Lambda + \Lambda + {^{A}Z}}$ model and in the Skyrme-Hartree-Fock
approach. Various ${\Lambda\Lambda}$ and $\Lambda$-nucleus or ${\Lambda N}$
potentials are used and the sensitivity of $\Delta B_{\Lambda\Lambda}$ to
the interactions is discussed. It is shown that in medium and heavy
${\Lambda\Lambda}$  hypernuclei, $\Delta B_{\Lambda\Lambda}$ is a linear
function of $r_{\Lambda}^{-3}$, where $r_\Lambda$ is rms radius of the
hyperon orbital.  It looks unlikely that it will be possible to extract
${\Lambda\Lambda}$ interaction from the double-$\Lambda$ hypernuclear
energies only, the additional information about the $\Lambda$-core
interaction, in particular, on $r_{\Lambda}$ is needed.
\end{abstract}

\newpage
\section{Introduction}

\par
Up to now, theoretical studies of double--$\Lambda$ hypernuclei were mainly
focused  on the light systems, for  only {$^{\ \;6}_{\Lambda\Lambda}$He},
{$^{\,10}_{\Lambda\Lambda}$Be} \cite{1}, and {$^{\,13}_{\Lambda\Lambda}$B}
\cite{2} have been identified experimentally. The key quantity of a
$\Lambda\Lambda$ hypernucleus is the bond energy of two $\Lambda$'s,
$\ \Delta B_{\Lambda\Lambda}\! =\! B_{\Lambda\Lambda}\! -\! 2B_{\Lambda}$,
where $B_{\Lambda\Lambda}$ is the separation energy of two ${\Lambda}$'s
from a ${^{A+2}_{\ \Lambda\Lambda}}Z$ hypernucleus and $B_{\Lambda}$ is
the ${\Lambda}$ separation energy in the corresponding single--${\Lambda}$
hypernucleus ${^{A+1}_{\ \ \;\Lambda}Z}$. The bond energy is conventionally
associated with the matrix element of ${\Lambda\Lambda}$ interaction in
the $^1S_0$ state.
\par
It is known that for light
${\Lambda\Lambda}$ hypernuclei, the bond energy is nearly constant,
$\Delta B_{\Lambda\Lambda}\! \sim \! 4.5$~MeV.
However, the respective information on ${\Lambda\Lambda}$ interaction
derived from this value
appears to be rather ambiguous due to substantial uncertainties in  the
shapes of both ${\Lambda\Lambda}$ and ${\Lambda A}$ potentials.
Using various ${\Lambda A}$ potentials that provide the same
$B_{\Lambda}$ values but different ${\Lambda}$-orbital radii
$r_{\Lambda}$, one derives very different
predictions for the ${\Lambda\Lambda}$ interaction
\cite{3,4,5}.
For example, the strength of ${\Lambda\Lambda}$ potentials of the same
shape extracted from the experimental $\Delta B_{\Lambda\Lambda}$ value
for {$^{\ \;6}_{\Lambda\Lambda}$He} using various $\Lambda\alpha$
potentials, differ by a factor of several times \cite{5}.
If the ${\Lambda\Lambda}$ potential is supposed to be purely attractive,
then the less is $r_{\Lambda}$ the larger is ${\Lambda\Lambda}$ attraction.
Nevertheless, one can believe, that the
information about the ${\Lambda\Lambda}$ interaction deduced from the
${\Lambda\Lambda}$ hypernuclear data, will become
less ambiguous if a larger number of ${\Lambda\Lambda}$ hypernuclei
differing essentially in masses, will be involved in the analysis.
\par
In  view of evolving $(K^-,K^+)$ experiments \cite{7,8}, we present
calculations of $\Delta B_{\Lambda\Lambda}$ for ${\Lambda\Lambda}$
hypernuclear ground states on a wide range of $A$. As far as we know, the
$A$-dependence of $\Delta B_{\Lambda\Lambda}$ has not been studied
systematically. The motivation  of our
study is two-fold. First, we investigate the general trend of the
$A$-dependence of $\Delta B_{\Lambda\Lambda}$. Next, we aim to examine how
much is the difference in the $\Delta B_{\Lambda\Lambda}$ predictions
obtained in distinct models using various $\Lambda\Lambda$
potentials, and whether it will
be possible to obtain the definite information about the ${\Lambda\Lambda}$
interaction from future experiments.
We calculate  the bond energies in two essentially
different approaches. In particular, we discuss below the results obtained
in the three-body model ${\Lambda + \Lambda +\; {^{A}Z}}$ with various
${\Lambda\Lambda}$ and ${\Lambda A}$ potentials in comparison with the
results of microscopic Hartree-Fock calculations with Skyrme-like
${\Lambda\Lambda}$, $\Lambda N$, and $NN$ interactions.
\section{Potentials and models}
Within the three-body approach, we treat a ${\Lambda\Lambda}$ hypernucleus
${^{A+2}_{\ \Lambda\Lambda}}Z$ as a ${\Lambda + \Lambda +\; {^{A}Z}}$ system
with the inert core ${^{A}Z}$. The conventional variational oscillator
basis expansion method has been used in calculations. The oscillator
frequency $\hbar\omega$ has been varied independently for each
hypernucleus to
minimize the ground state energy. Typically the complete $22\hbar\omega$
configuration space  has been used in calculations that provided the
convergence of the results. The convergence has been also checked by
the numerical location of the $S$-matrix pole corresponding to the ground
state using the technique of the harmonic oscillator representation of the
true three-body scattering theory \cite{SmSh} (see also \cite{12}).
\par
The $\Lambda A$ interaction has been described by the potentials proposed in
ref.~\cite{13} fitted to the single-$\Lambda$ hypernuclear spectra from
$(\pi^{+},K^+)$ reaction in the range of masses $A=8\div 88$ \cite{14}. The
potentials are also compatible with the recent data on $A=138$ and 207
hypernuclei \cite{15}. The first potential (hereafter referred to as C1)
is of the Woods-Saxon form:
\begin{equation}
{\rm C}1: \ \ \   \left\{
\begin{array}{l}
  V_{C1}(r)=V_{0}f(r), \ \ \ \ \ f(r)=\Bigl(1+e^{(r-c)/a}\Bigr)^{-1}; \\
\vphantom{\Biggr(\int_{0_{0_{0{0}}}} dx}
   V_0=-28\ {\rm MeV},\ \,
   c=\Bigl(1.128+0.439A^{-2/3}\Bigr)A^{1/3}\ {\rm fm},\\
   a=0.6\ {\rm fm}.  \label{1}
\end{array}
\right.
\end{equation}
\par
The second potential (C2) is the local potential of the $\rho^2$
density-de\-pen\-dent form \cite{13}:
\begin{equation}
{\rm C}2: \ \ \ \left\{
\begin{array}{l}
  V_{C2}(r)=V_{1}\rho (r) + V_{2}\rho^{2}(r), \ \ \ \ \
   \rho (r) = \rho_{0} f(r); \ \ \ \ \ \ \ \\
\vphantom{\Biggr(\int_{0_{0_{0{0}}}} dx}
    V_1=-340\ {\rm MeV},\ \,
    V_2=1087.5\ {\rm MeV},\\
     c=1.08A^{1/3}\ {\rm fm},\ \, a=0.54\ {\rm fm},  \label{2}
\end{array}
\right.
\end{equation}
where $\rho_{0} $ is a factor normalizing the nuclear density $\rho (r)$ to
$A$.
\par
Single-$\Lambda$ states generated by the potentials, have nearly the same
$B_{\Lambda}$ values but differ in the rms radii $r_{\Lambda}$.
\par
As for the ${\Lambda\Lambda}$ interaction, we used, firstly, the simplest
single-Gaussian potential with the range motivated by the two-pion
exchange model, that is widely used \cite{Ikeda} in double-$\Lambda$
hypernuclei calculations (hereafter referred to as $\Lambda 1$):
\begin{equation}
 \Lambda 1: \ \ \  \left\{
\begin{array}{l}
\vphantom{\Biggr(\int_{0_{0_{0{0}}}} dx}
 V_{\Lambda 1}(r)=V_{0}\exp\Bigl(-\frac{r^2}{r^{2}_{0}}\Bigr);
  \ \ \ \ \ \ \ \ \ \ \ \ \ \ \ \ \ \ \ \ \ \ \ \ \ \ \ \ \ \ \ \ \ \ \ \  \\
   r_0=1.034\ {\rm fm}.
\end{array}
\right.  \label{3}
\end{equation}
The second ${\Lambda\Lambda}$ potential ($\Lambda 2$) used is a two-Gaussian
paramet\-ri\-zation of ref.~\cite{16} with a repulsive core:
\begin{equation}
\Lambda 2: \ \ \ \left\{
\begin{array}{l}
\vphantom{\Biggr(\int_{0_{0_{0{0}}}} dx}
   V_{\Lambda 2}(r)=V_{1}\exp\Bigl(-\frac{r^2}{r^{2}_{1}}\Bigr)
 + V_{2}\exp\Bigl(-\frac{r^2}{r^{2}_{2}}\Bigr); \ \ \ \\
   V_{1} = 148\ {\rm MeV}, \ \, r_1= 0.82\ {\rm fm},\ \,
   r_2=1.2\ {\rm fm}. \ \ \ \ \ \ \ \ \
 \label{4}
\end{array} \right.
\end{equation}
\par
We have fitted the $\Lambda 1$ potential depth, $V_0$, and the attractive
component strength of the $\Lambda 2$ potential, $V_2$, to the bond energy
of {$^{\,13}_{\Lambda\Lambda}$B} ($\Delta B_{\Lambda\Lambda} = 4.8$~MeV
\cite{2,17,2st}) for each of the $\Lambda A$ potentials independently.
For boron, the $\Lambda A$ potentials considered give nearly the same
single-$\Lambda$ orbital radii $r_{\Lambda}$. Therefore, we have
obtained the values $V_0 = -62$~MeV and $V_2 = -107$~MeV adjusted to both
C1 and C2 potentials. The differences between the C1 and C2 radii
$r_{\Lambda}$ are pronounced in medium and heavy hypernuclei.
\par
We have performed also Hartree-Fock calculations of the bond energies with
Skyrme-like $NN$, $\Lambda N$, and $\Lambda\Lambda$ potentials. The
Skyrme-Hartree-Fock approach has been applied before to the single-$\Lambda$
hypernuclei, and Skyrme $\Lambda N$ potentials have been fitted to
single-$\Lambda$ hypernuclear spectra \cite{13,18,19,20}. This approach
has been extended to multi-$\Lambda$ systems in ref.~\cite{22}. Its
adaptation to double-$\Lambda$ hypernuclei is described elsewhere \cite{6}.
\par
$\Lambda\Lambda$ interaction in the ground states was described by a
simplified version of the Skyrme potential \cite{6},
\begin{eqnarray}
V(\vec{r}_1,\vec{r}_2) &=& \lambda_{0}\delta (\vec{r}_1-\vec{r}_2)
   +\frac{1}{2}\lambda_{1}\left[
{\vec{k'}}^{2}\delta (\vec{r}_1-\vec{r}_2)
+\delta (\vec{r}_1-\vec{r}_2){\vec{k}\vphantom{'}}^{\,2}\right],
\label{5}
\end{eqnarray}
with the conventional notations (see, e.g., \cite{18}).
the 
were used in calculations:  \begin{eqnarray} {\rm S}\Lambda\Lambda 1:\ \ &
  & \begin{array}{c} \lambda_0=-312.6\ {\rm MeV}\cdot{\rm fm}^3,\ \,
  \lambda_1=57.5\ {\rm MeV}\cdot{\rm fm}^5;\ \, 
         \end{array} \ \; \label{6}   \\[\medskipamount]
  {\rm S}\Lambda\Lambda 2:\ \ & &
     \begin{array}{c}
  \lambda_0=-437.7\ {\rm MeV}\cdot{\rm fm}^3,\ \,
  \lambda_1=240.7\ {\rm MeV}\cdot{\rm fm}^5;\ \, 
         \end{array} \ \;  \label{7}   \\[\medskipamount]
      {\rm S}\Lambda\Lambda 3:\ \ & &
     \begin{array}{c}
  \lambda_0=-831.8\ {\rm MeV}\cdot{\rm fm}^3,\ \,
  \lambda_1=922.9\ {\rm MeV}\cdot{\rm fm}^5.\ \, 
          \end{array} \ \;   \label{7'}
 \end{eqnarray}
The sets have been
fitted to the {$^{\,13}_{\Lambda\Lambda}$B} bond energy in ref.~\cite{1st}.
The S$\Lambda\Lambda$2 interaction is a Skyrme-like approximation
of the single-Gauss\-i\-an potential of the two-pion-exchange range
while the S$\Lambda\Lambda$1 interaction corresponds to a rather smaller
range. The S$\Lambda\Lambda$3 interaction possesses the largest range.
\par
The Hartree-Fock calculations also involved SkM$^*$ $NN$ interaction.
As for the $\Lambda N$ interaction, we mainly used YBZ5 potential
\cite{19}. This potential causes only a slight
polarization (distortion) of a nuclear core. The majority of the
$\Lambda N$ potentials fitted to single-$\Lambda$ hypernuclear spectra possess
this feature \cite{13,19,20} though  the case of the relatively strong
polarization cannot be excluded now.
%
%
\begin{figure}[bht]
\epsfverbosetrue
\epsfysize=10cm
\epsfbox{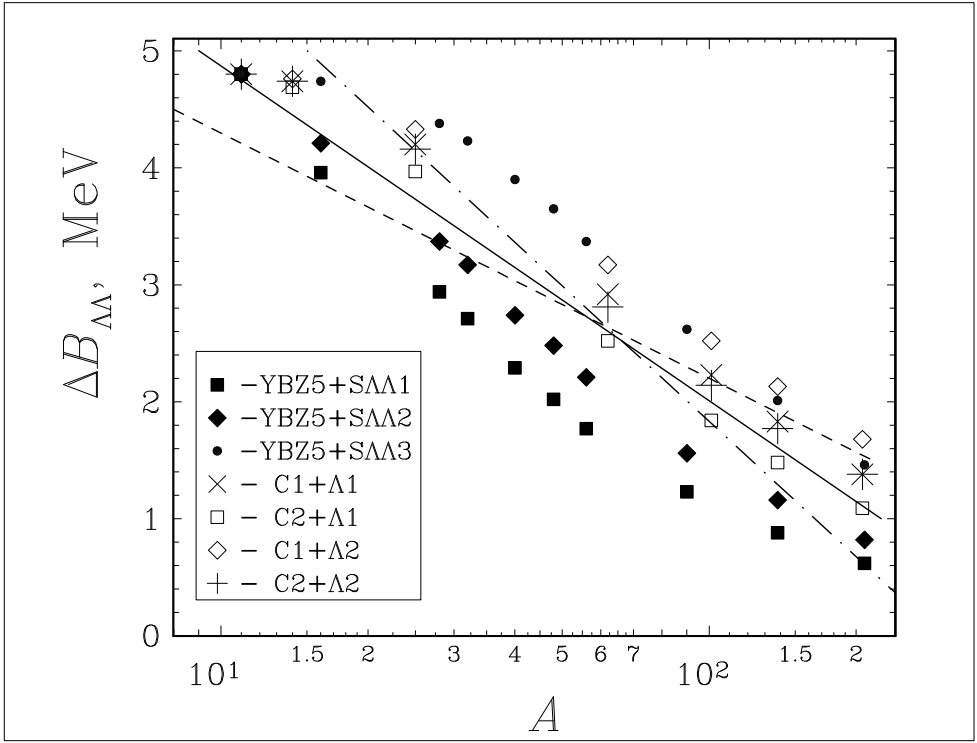}
\caption{$\Delta B_{\Lambda\Lambda}$
as a function of $A$. The results of the Skyrme-Hartree-Fock calculations
with $\Lambda N$ potential YBZ5 and various $\Lambda\Lambda$ interactions of
{$^{\,13}_{\Lambda\Lambda}$B}, {$^{\,18}_{\Lambda\Lambda}$O},
{$^{\,30}_{\Lambda\Lambda}$Si}, {$^{\,34}_{\Lambda\Lambda}$S},
{$^{\,42}_{\Lambda\Lambda}$Ca}, {$^{\,50}_{\Lambda\Lambda}$Ca},
{$^{\,58}_{\Lambda\Lambda}$Ni}, {$^{\,92}_{\Lambda\Lambda}$Zr},
{$^{140}_{\,\Lambda\Lambda}$La}, and {$^{210}_{\,\Lambda\Lambda}$Pb}
are presented by:
\filledboxes\ --- YBZ5+S$\Lambda\Lambda$1,
\filleddiamonds\ --- YBZ5+S$\Lambda\Lambda$2,
\protect\filledcircles~--- YBZ5+S$\Lambda\Lambda$3;
the results of the three-body
calculations with  various $\Lambda A$ and $\Lambda\Lambda$ interactions of
{$^{\,13}_{\Lambda\Lambda}$B}, {$^{\,16}_{\Lambda\Lambda}$C},
{$^{\,27}_{\Lambda\Lambda}$Na}, {$^{\,64}_{\Lambda\Lambda}$Co},
{$^{103}_{\,\Lambda\Lambda}$Ru}, {$^{140}_{\,\Lambda\Lambda}$La}, and
{$^{208}_{\,\Lambda\Lambda}$Hg} are presented by:
{$\times$} --- C1+$\Lambda$1, $\Box$ ---
C2+$\Lambda$1, $\Diamond$ --- C1+$\Lambda$2,
{$+$} --- C2+$\Lambda$2.
Solid, dashed, and dot-dashed lines are the plots of the functions
$\Delta B_{\Lambda\Lambda} = 15A^{-1/2}$~MeV,
$\Delta B_{\Lambda\Lambda} = 9A^{-1/3}$~MeV, and
$\Delta B_{\Lambda\Lambda} = 75A^{-1}$~MeV, respectively.}
\end{figure}
\section{Results and discussion}
The $A$-dependence of $\Delta B_{\Lambda\Lambda}$ is depicted on fig.~1.
It is seen that in general $\Delta B_{\Lambda\Lambda}$ decreases with $A$
increasing, although for small $A$ it is known to be practically
constant \cite{1,2,2st}. Attraction between hyperons become weaker just
because of the
increase of the mean ${\Lambda\Lambda}$ distance with $A$. In the
limit $A\to \infty$, evidently, $B_{\Lambda\Lambda} \to 2D_{\Lambda}$ and
$\Delta B_{\Lambda\Lambda} \to 0$, where $D_{\Lambda}$ is the
binding energy of the hyperon in the infinite nuclear matter.
\par
The differences between $\Delta B_{\Lambda\Lambda}$ values obtained in the
same approach  with different potentials are not very large, as well as ones
obtained by the two essentially different methods.
We have examined $\Delta B_{\Lambda\Lambda}$ in the three-body approach
using also other ${\Lambda\Lambda}$ potentials collected in ref.~\cite{5};
the results confirm the above conclusion.
For the same $\Lambda A$ potential, $\Delta B_{\Lambda\Lambda}$ increases
with the range of the ${\Lambda\Lambda}$ interaction. It is clearly
seen that only essential variations of potential parameters
(the $S\Lambda\Lambda 3$
interaction versus the $S\Lambda\Lambda 1$ and $S\Lambda\Lambda 2$ ones)
provides more or less considerable differences in the bond energies.
\par
It is seen that the $A$-dependence of
$\Delta B_{\Lambda\Lambda}$ in $\Lambda\Lambda$ hypernuclei is not very
sensitive to the details of the interactions and hence cannot provide
unambiguous information about the $\Lambda\Lambda$ potential strength and
shape.
\par
To get the estimate of the effects that the core polarization may have on
$\Delta B_{\Lambda\Lambda}$, we have performed also
the Skyrme-Hartree-Fock calculations with
two other $\Lambda N$ potentials: the SkSH1 \cite{20}
and the 3rd potential from \cite{18} (hereafter R3). The potentials cause the
strong polarization of the core and can be treated as two extreme examples of
what can happen with a nucleus when a $\Lambda$ hyperon is added to it:
the SkSH1 interaction results in contraction while the R3 in dilatation of
the core. Using strongly polarizing potentials one should renormalize
(sometimes considerably) the $\Lambda\Lambda$ interaction to reproduce
the boron bond energy \cite{6}, namely, the parameters $\lambda_0$ and
$\lambda_1$ of the S$\Lambda\Lambda$2 potential should be multiplied by
a factor of 0.33 in the case of SkSH1 interaction and by
a factor of 1.09 in the case of the R3 one.
\begin{figure}[tb]
\epsfverbosetrue
\epsfysize=10cm
\epsfbox{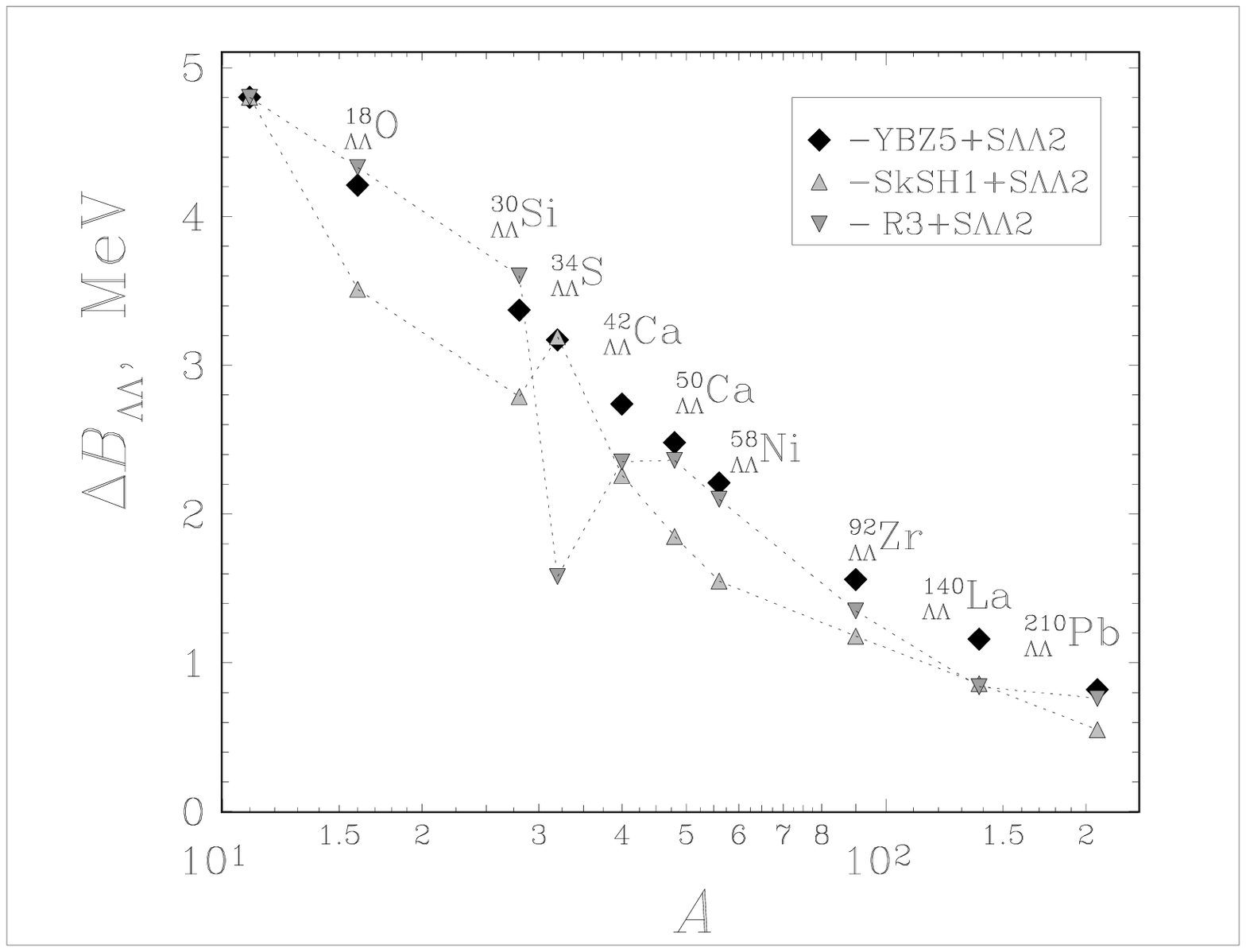}
\caption{The $A$-dependence of $\Delta B_{\Lambda\Lambda}$ obtained
in the Skyrme-Hartree-Fock approach with diluting R3
and contracting SkSH1 polarizing $\Lambda N$-interactions  in comparison
with the
results obtained with slightly polarizing YBZ5 $\Lambda N$-interaction:
\filleddiamonds\ --- YBZ5+S$\Lambda\Lambda$2,
\filledtriup\ --- SkSH1+S$\Lambda\Lambda$2,
\filledtridn\ --- R3+S$\Lambda\Lambda$2. Dotted lines connecting
the points are plotted to guide the eye. See fig.~1 for more details. }
\end{figure}
\par
The results of Skyrme-Hartree-Fock calculations with the strongly polarizing
$\Lambda N$ interactions are presented on fig.~2.
It is seen that the core polarization does not change the general trend of
the bound energy $A$-dependence. However, it results in
some irregularities of the $A$-dependence. The irregularities are associated
with some details of the $\Lambda N$ interactions.
For example, adding four $s$-nucleons to $^{\,30}_{\Lambda\Lambda}$Si
results in reduction of the $\Lambda$-orbital rms radius $r_{\Lambda}$
in $^{\,34}_{\Lambda\Lambda}$S
for the purely two-body SkSH1 potential. Therefore, the
$\Lambda\Lambda$ attraction increases. Contrary to it, in the case of the R3
interaction with strong
$\Lambda N$ repulsion, the hyperons in $^{\,34}_{\Lambda\Lambda}$S
are pushed out from the nuclear interior by the additional $s$ nucleons,
thus, $r_{\Lambda}$ increases and $\Delta B_{\Lambda\Lambda}$ decreases.
The reverse effect takes place when 8 $f$ nucleons are added to
$^{\,42}_{\Lambda\Lambda}$Ca and
$^{\,50}_{\Lambda\Lambda}$Ca forming $^{\,50}_{\Lambda\Lambda}$Ca and
$^{\,58}_{\Lambda\Lambda}$Ni, respectively. Note, however, that
$^{32}$S and $^{56}$Ni are not actually magic nuclei, so, the
$A$-dependence of the occupation numbers and, hence, of
$\Delta B_{\Lambda\Lambda}$ may be smoothed in more elaborate calculations.
\par
$\Delta B_{\Lambda\Lambda}$ appears to be nearly linear
function of $\log A$ for medium and heavy double-$\Lambda$ hypernuclei, i.e.,
$\Delta B_{\Lambda\Lambda}$ is proportional to $A^\alpha$.
To interpret this behavior, let us consider the following toy model.
Suppose
%
%
that the $\Lambda$-core potential $V_{\Lambda
A}(r)$ satisfies the scaling property, \begin{equation} V_{\Lambda
    A}(r)=V^{0}_{\Lambda A}f(r^{\prime}),\ \ \ \ r = r^{\prime} R_{A},
    \label{Scal}
\end{equation}
i.e., the strength of the potential, $V^{0}_{\Lambda A}$, is the same for
various heavy nuclei, and the $r$-dependence of $V_{\Lambda A}(r)$ is
described by some universal function $f(r^{\prime})$ of the scaled distance
$r^{\prime}=r/R_{A}$. The scaling property (\ref{Scal}) is exact, e.g., for
the
harmonic oscillator and rectangular well potentials.
\par
The potential scaling
(\ref{Scal}) implies  the scaling of the $\Lambda$ hyperon radial wave
function in the single-$\Lambda$ hypernuclei
${^{A+1}_{\ \Lambda\Lambda}}Z$,
\begin{equation}
        \phi_{\Lambda}^{A}(r) = N_{A}\varphi_{\Lambda}(r^{\prime}),
    \label{Norm}
\end{equation}
where $\varphi_{\Lambda}(r^{\prime})$ is the universal hyperon wave
function and the normalization constant $N_A \sim R_{A}^{-3/2}$.
We suppose that  $\Lambda\Lambda$ interaction, $V_{\Lambda\Lambda}(r)$,
is weak enough and does not perturb significantly the single-$\Lambda$
orbitals (\ref{Norm}) in the $\Lambda\Lambda$ hypernucleus
${^{A+2}_{\ \Lambda\Lambda}}Z$. If the range of
$V_{\Lambda\Lambda}(r)$ is small compared with the rms radius
$r_{\Lambda}$ of the orbital (\ref{Norm}), then for evaluation purposes we
can treat $V_{\Lambda\Lambda}(r)$ as a zero-range potential,
$V_{\Lambda\Lambda}(r)  = V_{\Lambda\Lambda}^{0} \delta(r)$, where
$\delta(r)$ is the Dirac $\delta$-function. In this case,
we immediately obtain, that
\begin{equation}
\Delta B_{\Lambda\Lambda} = |\langle V_{\Lambda\Lambda}(r) \rangle |=
V_{\Lambda\Lambda}^{0} R_{A}^{-3} \langle \varphi^{4}(r^{\prime}) \rangle,
\label{DB-sc}
\end{equation}
where the universal matrix element $\langle \varphi^{4}(r^{\prime}) \rangle$
does not depend on $A$.
\par
Supposing that $R_{A}=A^{1/3}$, i.e., that the range of the
$\Lambda A$ potential, $R_{\Lambda}^{0}$, is proportional to the radius
of the core ${^{A}_{\Lambda}Z}$, we get $\Delta B_{\Lambda\Lambda} \sim
A^{-1}$, while supposing that  $R_{A}=A^{1/6}$,
i.e., that $R_{\Lambda}^{0}$ is proportional to
the nucleon oscillator radius of the  oscillator shell model, we get
$\Delta B_{\Lambda\Lambda} \sim A^{-1/2}$. The  functions
$\Delta B_{\Lambda\Lambda} = const \cdot A^{-1/2}$ and
$\Delta B_{\Lambda\Lambda} = const \cdot A^{-1}$ are presented on fig.~1
by the solid and the dot-dashed lines, respectively, together with the
function $\Delta B_{\Lambda\Lambda} = const \cdot A^{-1/3}$ (the dashed
line). It is seen from the fig.~1, that the slope of the
$\Delta B_{\Lambda\Lambda}$ $A$-dependence obtained in Skyrme-Hartree-Fock
calculations for medium $\Lambda\Lambda$ hypernuclei is close to the one of
the function $\Delta B_{\Lambda\Lambda} = const \cdot A^{-1}$, while for
heavy $\Lambda\Lambda$ hypernuclei the slope is closer to the one of the
function $\Delta B_{\Lambda\Lambda} = const \cdot A^{-1/3}$ for the
$\Delta B_{\Lambda\Lambda}$ values obtained with the S$\Lambda\Lambda 1$ and
S$\Lambda\Lambda 2$ interactions and to the slope of the function
$\Delta B_{\Lambda\Lambda} = const \cdot A^{-1/2}$ for
$\Delta B_{\Lambda\Lambda}$ values obtained with the S$\Lambda\Lambda 3$
interaction. The slope of the $\Delta B_{\Lambda\Lambda}$ values obtained
in the three-body approach shows the behavior that is close to
$\Delta B_{\Lambda\Lambda} \sim A^{-1/2}$.
\par
However, the differences between
various $\Delta B_{\Lambda\Lambda}$ $A$-dependencies  are
not well-pronounced, and the main conclusion that we derive from the
fig.~1 is that very different models and very different interactions
yield more or less the same generally decreasing
$\Delta B_{\Lambda\Lambda}(A)$ behavior. At the same time, the effects of
the strong core polarization cause deviations from the general trend that are
seen from the fig.~2 to be unimportant in the heavy mass region.
\begin{figure}[tb]
\epsfverbosetrue
\epsfysize=10cm
\epsfbox{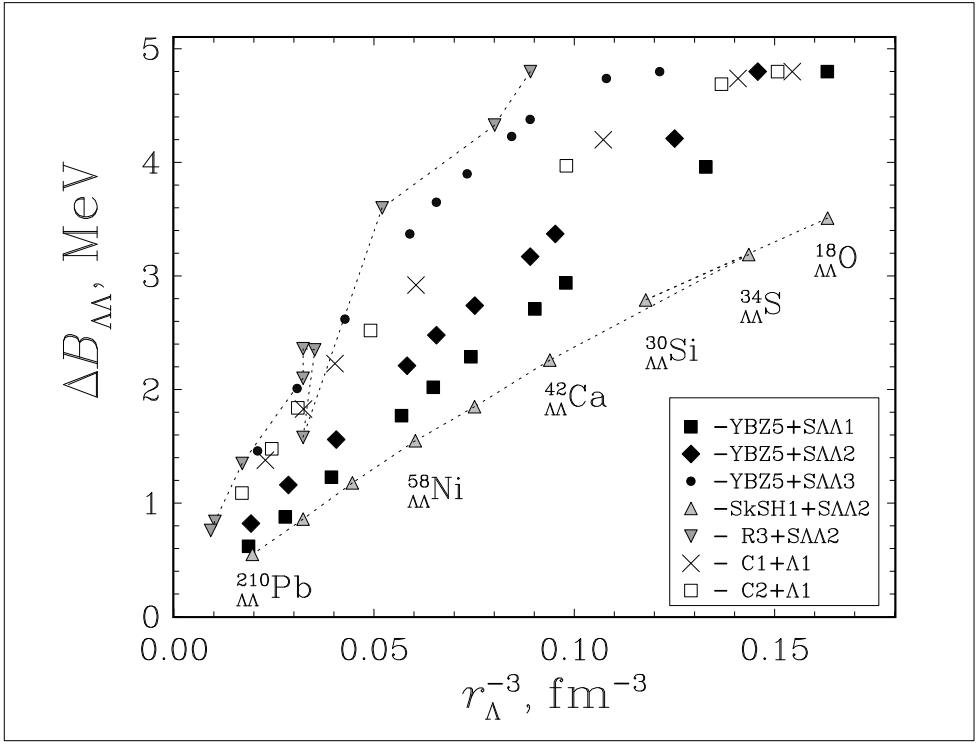}
\caption{The $\Delta B_{\Lambda\Lambda}$ scaling with $r_{\Lambda}^{-3}$
for various interactions.
See fig.~1 and 2 for details.}
\end{figure}
\par
As the rms radius of $\Lambda$ hyperons, $r_{\Lambda}$,
in  heavy $\Lambda\Lambda$ hypernuclei is proportional to $R_A$,
$r_{\Lambda} \sim R_A$, the scaling property (\ref{DB-sc}) means that
\begin{equation}
         \Delta B_{\Lambda\Lambda} \sim r_{\Lambda}^{-3}.
    \label{DB-r_L}
\end{equation}
The plot of $\Delta B_{\Lambda\Lambda}$
vs $r_{\Lambda}^{-3}$ is presented on fig.~3.  The scaling
(\ref{DB-r_L}) seems to be universal. The irregularities in the
$\Delta B_{\Lambda\Lambda}$ behavior obtained with the strongly polarizing
potentials are considerably smoothed in the $r_{\Lambda}^{-3}$-dependence.
In the case of SkSH1 interaction, the oscillation of the
$\Delta B_{\Lambda\Lambda}(A)$ dependence of
fig.~2 manifests itself  on fig.~3 only as an inverse order of the points
corresponding to $^{\,30}_{\Lambda\Lambda}$Si and
$^{\,34}_{\Lambda\Lambda}$S lying nearly on the same line with all other
points. Deviations from the single line of the points corresponding to R3
interaction are more pronounced, but they are much smaller than on fig.~2.
\par
In the heavy-mass (small $r_{\Lambda}^{-3}$) region the scaling
(\ref{DB-r_L}) appears to be very accurate. All the
$\Delta B_{\Lambda\Lambda}(r_{\Lambda}^{-3})$ dependencies obtained in our
calculations behave according to (\ref{DB-r_L}) in this region, i.e., the
$\Delta B_{\Lambda\Lambda}$ values for heavy $\Lambda\Lambda$
hypernuclei obtained in the
same way
lie on a straight line on the plane
$\Delta B_{\Lambda\Lambda}$ vs $r_{\Lambda}^{-3}$. The
deviation from the straight line in the
$\Delta B_{\Lambda\Lambda}(r_{\Lambda}^{-3})$ dependence in light and medium
mass $\Lambda\Lambda$ hypernuclei increases with the range of the
$\Lambda\Lambda$ interaction. Note, that the S$\Lambda\Lambda 1$ and
S$\Lambda\Lambda 2$ Skyrme-Hartree-Fock results presented on fig.~3 show the
linear behavior even in the light-mass region, while the
S$\Lambda\Lambda 3$  Skyrme-Hartree-Fock results that correspond to the
much larger range of the $\Lambda\Lambda$ interaction, deviate  from
the straight line  in the medium-mass region. The steepness of the
lines increases with the strength of the $\Lambda\Lambda$ interaction. Note,
that the points representing the results of the three-body calculations with
the same $\Lambda\Lambda$ interaction lie on the same curve.
\par
Concluding, we have calculated $A$-dependence of the bond
energy of $\Lambda$ hyperons in the $\Lambda\Lambda$ hypernuclear ground
states. This dependence is shown to be a generally decreasing function. The
$\Lambda\Lambda$ interaction cannot be extracted from the information about
the bond energies only even if they will be measured in a wide range of $A$.
On the other hand, the $\Delta B_{\Lambda\Lambda}$ $A$-dependence appears
to be  closely  connected with the
$A$-dependence of the rms radius of the hyperon orbital. The later has been
calculated for single-$\Lambda$ hypernuclei in various approaches
\cite{23,24,25,26}. However, no reliable information on $r_{\Lambda}$ has
been extracted from any available data up to now. It should be noted that
$r_\Lambda$'s calculated by us are not directly related with the rms radii
in single-$\Lambda$ hypernuclei due to some perturbation induced by the 2nd
hyperon. Nevertheless, this perturbation is not strong and does not change the
general trend. Therefore, the information on the rms radii of the hyperon
orbitals in $\Lambda$ hypernuclei is very important for deducing the
$\Lambda\Lambda$ interaction parameters from $\Lambda\Lambda$
hypernuclear data.
\par
Among the problems connected with the ${\Lambda\Lambda}$ hypernuclear
physics, the possible existence of $H$ dibaryon is of a great inte\-rest.
It is known,
that the existence of ${\Lambda\Lambda}$ hypernuclei restricts the
feasibility of long-living $H$ dibaryon, although does not rejects it
completely \cite{9,10}. On the other hand, up to now only three objects have
been observed that are conventionally identified as ${\Lambda\Lambda}$
hypernuclei, and the corresponding experimental data can be alternatively
reinterpreted  as $H$ hypernuclei \cite{11}. From this point of view, it
seems to be interesting to study the $A$-dependence of the bond energy for
$H$ hypernuclei, too, that can possibly serve for the discrimination of the
above alternatives.
\par
This work was supported in part by Russian Foundation for Fundamental
Investigations (RFFI) under the grant 94-02-04112 and by the Research
Program Russian Universities.

\end{document}